\documentclass[prd,twocolumn,showpacs,floatfix,amsmath,amssymb,nofootinbib]{revtex4}
\usepackage{bm}
\usepackage{graphicx}
\newcommand{\ud}{\mathrm{d}} 
 
\newcommand{\GF}{G_\mathrm{F}}
\newcommand{\eff}{\mathrm{eff}} 
\newcommand{\Hb}{\mathbf{H}}
\newcommand{\Htb}{\tilde{\mathbf{H}}}
\newcommand{\Sb}{\mathbf{S}}
\newcommand{\sB}{\mathbf{s}}

\newcommand{\ntot}{n_\nu^\mathrm{tot}}
\newcommand{\ft}{\tilde{f}}
\newcommand{\sgn}{\mathrm{sgn}}
\newcommand{\HV}{\Hb_\mathrm{V}}
\newcommand{\EC}{E_\mathrm{C}}
\newcommand{\wc}{\omega_\mathrm{pr}}
\newcommand{\wsync}{\omega_\mathrm{sync}}

\newcommand{\basef}[1]{\bm{\hat{\mathbf{e}}}^\mathrm{f}_{#1}}

\newcommand{\basev}[1]{\bm{\hat{\mathbf{e}}}^\mathrm{v}_{#1}}
\newcommand{\basetv}[1]{\bm{\hat{\tilde{\mathbf{e}}}}^\mathrm{v}_{#1}}
\newcommand{\thetav}{\theta_\mathrm{v}}

\newcommand{\myfigsep}{0.04 \textwidth}
\newcommand{\myfigwid}{0.48 \textwidth}

\begin{document}
\title{Simple Picture for Neutrino Flavor Transformation in Supernovae}
\newcommand*{\UCSD}{Department of Physics, %
University of California, San Diego, %
La Jolla, CA 92093-0319}
\newcommand*{\LANL}{Theoretical Division, Los Alamos National Laboratory, %
Los Alamos, NM 87545}
\newcommand*{\UMN}{School of Physics and Astronomy, %
University of Minnesota, Minneapolis, MN 55455}

\author{Huaiyu Duan}
\thanks{Present address: Institute for Nuclear Theory,
University of Washington, Seattle, WA 98195}
\email{hduan@phys.washington.edu}
\affiliation{\UCSD}
\author{George M.~Fuller}
\email{gfuller@ucsd.edu}
\affiliation{\UCSD}
% \author{J.~Carlson}
% \email{carlson@lanl.gov}
% \affiliation{\LANL}
\author{Yong-Zhong Qian}
\email{qian@physics.umn.edu}
\affiliation{\UMN}

\date{\today}

\begin{abstract}
We can understand many recently-discovered features of flavor evolution
in dense, self-coupled supernova neutrino and antineutrino systems
with a simple, physical scheme consisting of two quasi-static solutions. 
One solution
closely resembles the conventional, adiabatic single neutrino
Mikheyev-Smirnov-Wolfenstein (MSW) mechanism, 
in that neutrinos and antineutrinos
remain in mass eigenstates as they evolve in flavor space.
The other solution is analogous to the regular precession
of a gyroscopic pendulum in flavor space, and has been discussed
extensively in recent works.
Results of recent numerical studies are best explained with combinations
of these solutions in the following general scenario:
(1) Near the neutrino sphere, the MSW-like many-body solution obtains.
(2) Depending on neutrino vacuum mixing parameters, luminosities,
energy spectra, and the matter density profile,
collective flavor transformation in the nutation mode develops
and drives neutrinos away from the MSW-like evolution and
toward regular precession. (3) Neutrino and antineutrino
flavors roughly evolve according to the regular precession solution until
neutrino densities are low. In the late stage of the precession
solution, a stepwise swapping develops in the energy spectra
of $\nu_e$ and $\nu_\mu$/$\nu_\tau$.
We also discuss some subtle points regarding adiabaticity 
in flavor transformation in dense neutrino systems.
\end{abstract}


\maketitle

\section{Introduction}

Although it was well known that neutrino-neutrino forward scattering
could generate sizable flavor refractive indices 
in dense-neutrino environments
\cite{Fuller:1987aa,Notzold:1988kx,Pantaleone:1992xh,Sigl:1992fn,%
Fuller:1992aa,Qian:1993dg}, it was thought generally that
the large matter density near the supernova neutrino sphere
would suppress neutrino self-coupling effects.
However, it recently became  apparent that neutrino
self-coupling could alter drastically the way neutrinos
and antineutrinos evolve in flavor space in
the supernova environment \cite{Pastor:2002we,Balantekin:2004ug,Fuller:2005ae},
even for the small neutrino mass-squared differences
inferred from experiments \cite{Balantekin:2004ug,Fuller:2005ae}.

Assuming coherent neutrino flavor transformation and the efficacy
of a mean field approach \cite{Friedland:2003eh,Balantekin:2006tg}, and
using the magnetic spin analogy for $2\times2$ neutrino oscillations
and the technique of the ``corotating frame'' (discussed later
in the text), Ref.~\cite{Duan:2005cp} was able to show that
neutrinos can experience collective flavor transformation
even in the presence of a dominant matter field.
The first large-scale numerical simulations of neutrino flavor
transformation in the late-time supernova environment with
correlated neutrino trajectories were discussed in 
Refs.~\cite{Duan:2006an,Duan:2006jv}. The results of these simulations 
clearly revealed the collective nature of neutrino oscillations
in supernovae. These effects are not explainable by the conventional
Mikheyev-Smirnov-Wolfenstein (MSW) effect
\cite{Wolfenstein:1977ue,Wolfenstein:1979ni,Mikheyev:1985aa}.

The numerical simulations \cite{Duan:2006an,Duan:2006jv} showed 
that neutrinos of all energies and propagating
along all trajectories can experience rapid flavor oscillations 
on short time/distance scales in the late-time supernova hot-bubble region.
This type of collective flavor transformation was actually
first discussed in the context of the early universe 
\cite{Kostelecky:1993dm,Kostelecky:1994dt,Samuel:1995ri}.
Ref.~\cite{Hannestad:2006nj} showed that a uniform mono-energetic 
neutrino gas initially in pure $\nu_e$ and $\bar\nu_e$ states is
equivalent to a gyroscopic pendulum in flavor space,
and that the rapid collective neutrino flavor transformation
 in supernovae and in the early universe corresponds to the
nutation of this flavor pendulum.

The simulations discussed in Refs.~\cite{Duan:2006an,Duan:2006jv}
also revealed a stunning feature of collective neutrino flavor transformation:
$\nu_e$'s appear to swap
their energy spectra with $\nu_{\mu/\tau}$'s at energies below or above 
(depending on the neutrino mass hierarchy) 
a transition energy $E_\mathrm{C}$.
It was suggested in Ref.~\cite{Duan:2006an} that this
stepwise swapping in neutrino energy spectra is related
to a collective mode in which the relative phases of
the two flavor components of neutrinos oscillate at the same rate.
Ref.~\cite{Duan:2007mv} argued that this collective mode is related to the
regular precession of a flavor pendulum.
This was done by solving the equations for
regular precession mode of a  gas of
 mono-energetic neutrinos  initially in pure $\nu_e$ 
and $\bar\nu_e$ states. Following the suggestion
 proposed in Ref.~\cite{Duan:2006an},
Ref.~\cite{Duan:2007mv} further elucidates why
a stepwise swapping in neutrino energy spectra can result from
neutrinos evolving in a collective precession mode. 
Ref.~\cite{Raffelt:2007cb} found a way to obtain the regular precession solution
for  neutrino gases with any given neutrino number
density and energy spectra. 
Ref.~\cite{Raffelt:2007cb} also pointed out that the 
transition energy $E_\mathrm{C}$
was determined by the conserved lepton number in the regular
precession solution.  This solution 
was termed an ``adiabatic solution'' in
Ref.~\cite{Raffelt:2007cb}. However, we will show later in this paper
that the flavor evolution in the regular precession solution is actually
not adiabatic.

Inspired by the method that Ref.~\cite{Raffelt:2007cb} used to
find the regular precession solution, we show that a quasi-static
solution for given electron density and neutrino number density can
be derived in a similar way. This solution is very much like the
conventional adiabatic MSW solution, except that the background
neutrinos contribute
 an additional refractive index to a given propagating neutrino.
We calculate this MSW-like solution as well as the regular precession
solution under the same conditions as the single-angle numerical simulations 
discussed in Refs.~\cite{Duan:2006an,Duan:2006jv}. The 
comparison between these solutions and the numerical results
renders a simple picture 
for neutrino flavor transformation in supernovae: Neutrinos
initially follow the MSW-like solution near the neutrino sphere
before being driven away from this solution by collective flavor
transformation in the nutation mode. Subsequently, neutrinos roughly
follow the regular precession solution with some nutation. 
As the neutrino and antineutrino densities decrease, 
 stepwise swapping appears in the neutrino energy spectra.

This paper is organized as follows. In Sec.~\ref{sec:adiabatic-sol}
we derive the MSW-like solution and discuss the conditions 
under which flavor
evolution in this scenario is adiabatic. In Sec.~\ref{sec:prec-sol}
we briefly recapitulate the derivation of the regular
precession solution and discuss the effects of the matter field
on this solution. In Sec.~\ref{sec:supernova} we compare the MSW-like
solution and the regular precession solution to the numerical results
and discuss the origin of stepwise swapping in neutrino energy spectra. We also
elaborate on the adiabaticity of collective
neutrino flavor evolution. In Sec.~\ref{sec:conclusions} we give
our conclusions.

\section{MSW-like solution  for dense neutrino gases%
\label{sec:adiabatic-sol}}
\subsection{Equations of motion%
\label{sec:eom}}

Here we briefly recapitulate a discussion of
the equations of motion (e.o.m.)
for neutrino flavor transformation. We use the
notation of neutrino flavor isospin (NFIS) 
(see Ref.~\cite{Duan:2005cp} for detailed discussions).
A NFIS is defined as the expectation value of the Pauli spin
operator of the flavor wavefunction for a neutrino or antineutrino.
The flavor basis wavefunction for a neutrino in the $2\times2$ mixing scheme is
\begin{equation}
\psi_\nu\equiv
\begin{pmatrix}
a_{\nu_e}\\
a_{\nu_\tau}
\end{pmatrix},
\end{equation}
where $a_{\nu_e}$ and $a_{\nu_\tau}$ are the amplitudes for the neutrino
to be  in $\nu_e$ and another flavor state, say $\nu_\tau$, respectively. 
The corresponding NFIS in the flavor basis is 
\begin{equation}
\sB_\nu\equiv\psi_\nu^\dagger\frac{\bm{\sigma}}{2}\psi_\nu
=\frac{1}{2}
\begin{pmatrix}
2 \mathrm{Re} (a_{\nu_e}^* a_{\nu_\tau})\\
2 \mathrm{Im} (a_{\nu_e}^* a_{\nu_\tau}) \\
|a_{\nu_e}|^2-|a_{\nu_\tau}|^2
\end{pmatrix}.
\label{eq:snu}
\end{equation}
The flavor basis wavefunction for an antineutrino
is defined as
\begin{equation}
\psi_{\bar\nu}\equiv
\begin{pmatrix}
-a_{\bar\nu_\tau}\\
a_{\bar\nu_e}
\end{pmatrix},
\end{equation}
where $a_{\bar\nu_e}$ and $a_{\bar\nu_\tau}$ are the amplitudes for the 
antineutrino to be  $\bar\nu_e$ and  $\bar\nu_\tau$, respectively.
The corresponding NFIS in the flavor basis is 
\begin{equation}
\sB_{\bar\nu}\equiv \psi_{\bar\nu}^\dagger\frac{\bm{\sigma}}{2}\psi_{\bar\nu}
=-\frac{1}{2}
\begin{pmatrix}
2 \mathrm{Re} (a_{\bar\nu_e} a_{\bar\nu_\tau}^*)\\
2 \mathrm{Im} (a_{\bar\nu_e} a_{\bar\nu_\tau}^*) \\
|a_{\bar\nu_e}|^2-|a_{\bar\nu_\tau}|^2
\end{pmatrix}.
\label{eq:sanu}
\end{equation}
As one will see later, the special definition of the
flavor basis wavefunction and NFIS for the antineutrino allows one
to write the e.o.m.~for both neutrinos and antineutrinos
in a unified manner.

The flavor evolution of a neutrino is represented by the precession
of the corresponding NFIS around an effective external field $\Hb$.
The  e.o.m.~of  NFIS $\sB_i$ can be written in a
form similar to that for a magnetic spin:
\begin{equation}
\frac{\ud}{\ud t} \sB_i = \sB_i\times\Hb_i
\label{eq:eom}
\end{equation}
where subscript $i$ denotes the initial physical state 
(flavor species and momentum), or ``mode'', of the neutrino/antineutrino.
As $\sB_i$ precesses around $\Hb_i$,
the projection of $\sB_i$ on $\basef{z}$ oscillates.
This represents
variation in the flavor content of the neutrino.
Here $\basef{x}$, $\basef{y}$ and $\basef{z}$ 
are the unit vectors in the flavor basis.

In the absence of ordinary matter and other neutrinos
(in vacuum), the effective
field experienced by NFIS $\sB_i$ is 
\begin{equation}
\Hb_i = \omega_i \HV,
\label{eq:Hi-vac}
\end{equation}
where the vacuum field is
\begin{equation}
\HV \equiv -\basef{x}\sin2\thetav + \basef{z}\cos2\thetav,
\end{equation}
and $\omega_i$ is
\begin{equation}
\omega_i \equiv \pm\frac{\delta m^2}{2 E_i}.
\label{eq:muV-def}
\end{equation}
In this notation $\thetav\in(0,\pi/2)$ is the 
effective $2\times2$ vacuum mixing angle,
$\delta m^2>0$ is the mass-squared difference between 
two neutrino vacuum mass eigenstates, and $E_i$
is the energy of the neutrino. 
The plus sign in Eq.~\eqref{eq:muV-def} is for neutrinos, and the
minus sign is for antineutrinos.
Throughout this paper we will assume that $\sin2\thetav=\sin2\theta_{13}\ll1$.
In this convention one has $\thetav\ll1$ for the normal neutrino
mass hierarchy and $\thetav\simeq\pi/2$ for the inverted one.
Eqs.~\eqref{eq:eom} and \eqref{eq:Hi-vac} for NFIS $\sB_i$
can be  pictured as the precession 
of $\sB_i$ around $\HV$ with angular velocity $\omega_i$.

In the presence of ordinary matter, the forward scattering
of neutrinos on electrons induces different refractive indices
for neutrinos of electron flavor from those for other flavors. As a result, 
in the presence of matter,
NFIS $\sB_i$ will also tend to precess around the vector field
\begin{equation}
\Hb_e \equiv -\sqrt{2}\GF n_e\basef{z}.
\end{equation}
Here $\GF$ is the Fermi constant and
$n_e$ is the net electron number density.

For dense neutrino gases,  forward neutrino-neutrino scattering
couples the flavor evolution of different neutrino modes.
With all the contributions taken into account,
the effective field experienced by NFIS $\sB_i$ is
\begin{equation}
\Hb_i = \omega_i\HV+\Hb_e+\mu_\nu \sum_j n_j \sB_j,
\label{eq:Hi}
\end{equation}
where 
\begin{equation}
\mu_\nu\equiv-2\sqrt{2}\GF
\end{equation}
is the coupling coefficient of two neutrino modes
for isotropic neutrino gases,
and $n_j$ is the population (number density) of neutrino mode $j$.
We define the normalized distribution function as
\begin{equation}
\ft_i\equiv \frac{n_i}{\ntot},
\end{equation}
where
\begin{equation}
\ntot \equiv \sum_i n_i
\end{equation}
is the total neutrino number density.

\subsection{MSW-like solution%
\label{sec:adiabatic-sol-der}}

In the NFIS notation, the instantaneous light (heavy)  mass eigenstate
of a neutrino is represented by a NFIS completely aligned (anti-aligned)
with its effective field. (Note that the effective energy of NFIS $\sB_i$
is $\varepsilon^\eff_i=-\sB_i\cdot\Hb_i$.) 
If the flavor evolution of neutrinos is fully
adiabatic, a neutrino initially in the light (heavy) mass eigenstate
will stay in the same instantaneous 
mass eigenstate, and a NFIS initially aligned
(anti-aligned) with its effective field will stay aligned (anti-aligned)
with the effective field. 
In such a limit one has
\begin{equation}
\sB_i = \epsilon_i \frac{\Hb_i}{2H_i},
\label{eq:alignment}
\end{equation}
where the alignment factor $\epsilon_i=+1$ ($-1$) if 
NFIS $\sB_i$ is aligned (anti-aligned)
with its effective field $\Hb_i$.

It is convenient to work in the vacuum mass basis in which
the unit vectors $\basev{x(y,z)}$ are related to $\basef{x(y,z)}$ by
\begin{subequations}
\begin{align}
\basev{x}&=\basef{x}\cos2\thetav+\basef{z}\sin2\thetav,\\
\basev{y}&=\basef{y},\\
\basev{z}&=\HV=-\basef{x}\sin2\thetav+\basef{z}\cos2\thetav.
\end{align}
\end{subequations}
Using Eq.~\eqref{eq:Hi}
we can express Eq.~\eqref{eq:alignment} in the explicit component form
in the vacuum mass basis:
\begin{subequations}
\label{eq:sixyz}
\begin{align}
s_{i,x} &= \frac{\epsilon_i}{2H_i}(H_{e,x}+\mu_\nu S_x),
\label{eq:six}\\
s_{i,y} &= \frac{\epsilon_i}{2H_i}\mu_\nu S_y,
\label{eq:siy}\\
s_{i,z} &= \frac{\epsilon_i}{2H_i}(\omega_i+H_{e,z}+\mu_\nu S_z)
\label{eq:siz}.
\end{align}
\end{subequations}
Here $S_{x(y,z)}$ are the components of the total NFIS
\begin{equation}
\Sb \equiv \sum_i n_i \sB_i=\ntot \sum_i \ft_i \sB_i
\label{eq:Stot}
\end{equation}
in the same basis, and
\begin{equation}
H_i = \sqrt{(H_{e,x}+\mu_\nu S_x)^2+(\mu_\nu S_y)^2
+(\omega_i+H_{e,z}+\mu_\nu S_z)^2}.
\end{equation}
Summing Eq.~\eqref{eq:sixyz} over index $i$ with 
weight $n_i=\ntot\ft_i$ we obtain
\begin{subequations}
\begin{align}
S_x &= \frac{\ntot}{2}(H_{e,x}+\mu_\nu S_x)\sum_i\frac{\epsilon_i\ft_i}{H_i},
\label{eq:six-sum}\\
S_y &= \frac{\ntot}{2}\mu_\nu S_y\sum_i\frac{\epsilon_i\ft_i}{H_i},
\label{eq:siy-sum}\\
S_z &= \frac{\ntot}{2}\sum_i
\frac{\epsilon_i\ft_i}{H_i}(\omega_i+H_{e,z}+\mu_\nu S_z).
\end{align}
\end{subequations}
Eqs.~\eqref{eq:six-sum} and \eqref{eq:siy-sum} imply $S_y=0$ if $H_e\neq0$. 
In this case, and for any given $\ntot$ and $n_e$, 
one can solve for $S_x$ and $S_z$ using the equations
%\begin{widetext}
\begin{subequations}
\label{eq:Sxz}
\begin{align}
S_x &= \frac{\ntot}{2}(H_{e,x}+\mu_\nu S_x)
\nonumber\\
&\quad\times
\sum_i\frac{\epsilon_i\ft_i}%
{\sqrt{(H_{e,x}+\mu_\nu S_x)^2+(\omega_i+H_{e,z}+\mu_\nu S_z)^2}},
\label{eq:Sx}\\
S_z &= \frac{\ntot}{2}
%\nonumber\\
%&\quad\times
\sum_i\frac{\epsilon_i\ft_i(\omega_i+H_{e,z}+\mu_\nu S_z)}%
{\sqrt{(H_{e,x}+\mu_\nu S_x)^2+(\omega_i+H_{e,z}+\mu_\nu S_z)^2}}.
\label{eq:Sz}
\end{align}
\end{subequations}
%\end{widetext}
The components of NFIS $\sB_i$
can then be obtained from Eq.~\eqref{eq:sixyz}.
We note that the counterpart of Eq.~\eqref{eq:Sxz}
in the flavor basis was discussed in connection
with adiabatic flavor transformation of supernova
neutrinos in Ref.~\cite{Qian:1994wh}.

We also note that the vanishing of the $z$-component of $\Sb$ 
in the flavor basis, \textit{i.e.},
\begin{equation}
S_z^\mathrm{f}=S_z\cos2\thetav+S_x\sin2\thetav=0,
\end{equation}
corresponds to an MSW-like resonance at which the total NFIS
represents a maximally mixed state. Using Eqs.~\eqref{eq:Sx} and \eqref{eq:Sz}
we rewrite the above equation as
\begin{equation}
\left(\sum_i\frac{\epsilon_i\tilde f_i}{H_i}\omega_i\right)\cos2\thetav
+\left(\sum_i\frac{\epsilon_i\tilde f_i}{H_i}\right)H_{e,z}^\mathrm{f}=0,
\label{eq:MSW-resonance}
\end{equation}
where $H_{e,z}^\mathrm{f}=-\sqrt{2}G_\mathrm{F}n_e$.
In the limit $\ntot\rightarrow\infty$,
all NFIS's are either
aligned or anti-aligned with the total NFIS:
\begin{equation}
\sB_i=\epsilon_i\frac{\Hb_i}{2H_i}
\simeq-\epsilon_i\frac{\Sb}{2S}
\end{equation}
(note that $\mu_\nu<0$). In the same limit,
Eq.~\eqref{eq:Sx} simply gives a normalization condition
\begin{equation}
\frac{\mu_\nu\ntot}{2}\sum_i\frac{\epsilon_i\ft_i}{H_i}
\simeq \frac{\ntot}{2S}\sum_i (-\epsilon_i)\ft_i
\simeq1,
\end{equation}
and Eq.~\eqref{eq:MSW-resonance} reduces to
\begin{equation}
\wsync\cos2\thetav+H_{e,z}^\mathrm{f}\simeq0,
\label{eq:MSW-resonance2}
\end{equation}
where
\begin{equation}
\wsync\equiv\frac{1}{S^2}\sum_i\omega_in_i\sB_i\cdot\Sb
\simeq\frac{n_\nu^\mathrm{tot}}{2S}\sum_i(-\epsilon_i)\tilde f_i\omega_i
\end{equation}
is the synchronization frequency \cite{Pastor:2001iu}.
Eq.~\eqref{eq:MSW-resonance2} implies
that all neutrinos go through an MSW-like resonance
at approximately 
 the radius where a neutrino with energy $E=\delta m^2/2|\wsync|$
would for a conventional, single-neutrino MSW resonance.

Eqs.~\eqref{eq:six-sum} and \eqref{eq:siy-sum} 
are not independent 
if $H_e=0$. Assuming $\Sb$ is  static with given $\ntot$, one
can still choose a new coordinate system in which $S_y$ vanishes.
Naively, it might seem practical to solve for $S_x$ and $S_z$ using
Eqs.~\eqref{eq:Sx} and \eqref{eq:Sz}. However, 
from Eqs.~\eqref{eq:eom} and \eqref{eq:Hi} one can show that
the lepton number
\begin{equation}
\mathcal{L} = 2\sum_i \ft_i \sB_i\cdot\HV = \frac{2S_z}{\ntot}
\label{eq:L}
\end{equation}
is conserved for $H_e=0$ \cite{Hannestad:2006nj}.
This is true even if $\ntot$ is a function of time \cite{Duan:2007mv}.
Therefore, $S_x$ and $S_z$ are over constrained by 
Eqs.~\eqref{eq:Sx}, \eqref{eq:Sz} and \eqref{eq:L}.
In this case, a new unknown is
expected.
In fact, a regular precession solution exists in the absence of
the matter field. In such a solution, $\Sb$ precesses around
$\HV$ and the precession angular velocity $\wc$ is the new unknown variable
to be found (see Sec.~\ref{sec:prec-sol}).

\subsection{Adiabatic condition%
\label{sec:adiabatic-cond}}

The MSW-like solution inherent in Eqs.~\eqref{eq:Sx} and \eqref{eq:Sz} is
a quasi-static solution for any given $n_e$ and $\ntot$.
In this solution, as $\Hb_i$ changes with varying $n_e$ and $\ntot$,
$\sB_i$ can lag behind. In other words, as the ensemble of neutrinos evolves,
misalignment between $\sB_i$ and $\Hb_i$ can develop.
Consequently, $\sB_i$ will tend to
 precess around $\Hb_i$ with angular speed $H_i$. Therefore, 
in order for  the flavor evolution of neutrino mode $i$ to follow
the MSW-like solution, one must have
\begin{equation}
\gamma_i \equiv H_i^{-1} \left|\frac{\ud  \vartheta_i}{\ud t}\right| \ll1,
\label{eq:adiabatic-cond}
\end{equation}
where $\vartheta_i$ is the angle between $\Hb_i$ 
and $\HV$ in the MSW-like solution. Eq.~\eqref{eq:adiabatic-cond}
is  the condition for the MSW-like solution to be adiabatic.

We note that the MSW-like solution becomes more adiabatic with larger
neutrino densities. This is because $H_i$
is the energy gap between the instantaneous light and heavy mass eigenstates
of neutrino mode $i$, and it increases with $\ntot$.

We also note that Eq.~\eqref{eq:adiabatic-cond} only gives
the necessary condition for the neutrino system to adiabatically follow the
MSW-like solution. This is because in deriving Eq.~\eqref{eq:adiabatic-cond}
we have assumed that $\Hb_i$ is described by 
the MSW-like solution in the first place.
This is true for the conventional MSW solution where $\ntot=0$.
If $\ntot$ is large, however, all NFIS's can move in a collective
manner, even in the presence of the matter field \cite{Duan:2005cp}.
As a result, the total NFIS $\Sb$ and the effective field 
$\Hb_i=\omega_i\HV+\Hb_e+\mu_\nu\Sb$ may not follow the MSW-like solution 
at all. We will further elaborate
on this point in Sec.~\ref{sec:adiabaticity}.

\section{Regular precession solution for dense neutrino gases%
\label{sec:prec-sol}}
\subsection{Regular precession solution%
\label{sec:prec-sol-der}}

In the absence of a matter field ($H_e=0$), the e.o.m.~of NFIS's
possess  cylindrical symmetry around $\HV$. A solution with the 
same symmetry is expected to exist for any given $\ntot$. In this
solution, which
 we term the ``regular precession solution'', all NFIS's must precess
steadily around $\HV$ and
\begin{equation}
\frac{\ud}{\ud t}\sB_i = \sB_i \times \wc\HV.
\label{eq:eom-regular-prec}
\end{equation}
For a  gas of mono-energetic neutrinos
initially in pure $\nu_e$ and $\bar\nu_e$ states, this represents
the regular precession of the gyroscopic pendulum in  flavor space
\cite{Duan:2007mv}.

Combining Eqs.~\eqref{eq:eom}, \eqref{eq:Hi} and \eqref{eq:eom-regular-prec},
we have
\begin{equation}
\sB_i\times[(\omega_i-\wc)\HV+\mu_\nu\Sb] = 0,
\label{eq:prec}
\end{equation}
which means that $\sB_i$ is either aligned or anti-aligned with
\begin{equation}
\Htb_i\equiv(\omega_i-\wc)\HV+\mu_\nu\Sb.
\label{eq:Hti}
\end{equation}
We note that $\Htb_i$ is the effective field experienced by NFIS $\sB_i$
in a frame whose
unit vectors $\basetv{x(y,z)}$ 
 rotate in the static frame according to
\begin{equation}
\frac{\ud}{\ud t}\basetv{x(y,z)}=\basetv{x(y,z)}\times\wc\HV.
\end{equation}
Therefore, the regular precession solution can be derived
by means similar to those used to get the MSW-like solution
\cite{Raffelt:2007cb}.

Eq.~\eqref{eq:Hti} shows that the effective fields
for all NFIS's and, therefore, all NFIS's themselves, reside in 
a static plane in the corotating frame which is spanned by
$\HV$ and $\Sb$. Without loss of
generality, we assume this 
corotating plane to be the $\basetv{x}$--$\basetv{z}$ plane,
and write out the components of NFIS $\sB_i$:
\begin{subequations}
\label{eq:si}
\begin{align}
s_{i,x} =& \frac{\epsilon_i}{2} \frac{\mu_\nu S_x}%
{\sqrt{(\omega_i-\wc+\mu_\nu S_z)^2+(\mu_\nu S_x)^2}},
\label{eq:six-prec}\\
s_{i,z} =& \frac{\epsilon_i}{2} \frac{\omega_i-\wc+\mu_\nu S_z}%
{\sqrt{(\omega_i-\wc+\mu_\nu S_z)^2+(\mu_\nu S_x)^2}},
\label{eq:siz-prec}
\end{align}
\end{subequations}
where subscripts $x$ and $z$ indicate the projections of the vectors onto
$\basetv{x}$ and $\basetv{z}$, respectively, and 
the alignment factor is $\epsilon_i=+1$ ($-1$) for
$\sB_i$  aligned (anti-aligned) with $\Htb_i$.
We sum Eq.~\eqref{eq:six-prec} with weight $\ntot\ft_i$ and obtain
\begin{equation}
1 = \frac{\mu_\nu\ntot}{2} \sum_i \frac{\epsilon_i \ft_i}%
{\sqrt{(\omega_i-\wc+\mu_\nu S_z)^2+(\mu_\nu S_x)^2}}.
\label{eq:unity}
\end{equation}
Summing Eq.~\eqref{eq:siz-prec}
with weight $\ntot\ft_i$ and
using Eq.~\eqref{eq:unity}, we obtain
\begin{equation}
\wc = \frac{\mu_\nu\ntot}{2} \sum_i \frac{\epsilon_i\omega_i \ft_i}%
{\sqrt{(\omega_i-\wc+\mu_\nu S_z)^2+(\mu_\nu S_x)^2}}.
\label{eq:wc}
\end{equation}
For any given values of total neutrino  number density $\ntot$ and 
lepton number $\mathcal{L}$, 
 Eq.~\eqref{eq:L} specifies $S_z$. One can then obtain
$S_x$ and $\wc$ from Eqs.~\eqref{eq:unity} and \eqref{eq:wc}.
The components of each NFIS in the corotating frame are
determined by Eq.~\eqref{eq:si}.

\subsection{Pseudo-regular precession condition%
\label{sec:pseudo-rp-cond}}

Because the angle between $\sB_i$ and $\HV$ varies
with $\ntot$, it is not possible for $\sB_i$ to stay in the
regular precession solution. In this case, $\sB_i$ will tend to
precess around $\Htb_i$ in the corotating frame. This precession
corresponds to nutation  in the static frame.
Ref.~\cite{Duan:2007mv} has argued that the regular precession solution
can be an excellent approximation to the actual evolution of the system
if the inverse of the
nutation time scale is much larger than $|\ud \vartheta_i/\ud t|$,
where $\vartheta_i$ is the angle between $\sB_i$ and $\HV$
in the regular precession solution.
As $\sB_i$ is parallel to $\Htb_i$, $\vartheta_i$
is equivalent to the angle between $\Htb_i$ and $\HV$ for the solution.
 If this is the case, the amplitude
of the nutation will be small. This behavior is similar to
the pseudo-regular precession of a gyroscope.

We note that the time scale of the nutation
of $\sB_i$ in the static frame is the same as the period of the precession
of $\sB_i$ around $\Htb_i$ in the corotating frame,
which is $\tilde{H}_i$. Therefore, the
condition for NFIS $\sB_i$ to be in the pseudo-regular precession is
\begin{equation}
\tilde{\gamma}_i \equiv \tilde{H}_i^{-1}
\left|\frac{\ud  \vartheta_i}{\ud t}\right| \ll1.
\label{eq:pseudo-rp-cond}
\end{equation}

We note that Eq.~\eqref{eq:pseudo-rp-cond} only gives the necessary
condition for neutrino systems to follow the regular precession solution. 
This is because in deriving Eq.~\eqref{eq:pseudo-rp-cond} we have assumed that
$\Htb_i$ follows the regular precession solution and rotates
much slower than does $\sB_i$. This can be true
if the nutation of all NFIS's are not correlated.
In collective flavor transformation, however, the motion
of the total NFIS $\Sb$ and, therefore, $\Htb_i$ is correlated
with individual NFIS $\sB_i$. As a result, $\Htb_i$ moves
at a rate comparable to that of $\sB_i$, and the flavor evolution
may not exactly follow the regular precession solution.

The pseudo-regular precession condition
in Eq.~\eqref{eq:pseudo-rp-cond} was first proposed
in Ref.~\cite{Raffelt:2007cb} as an ``adiabatic condition''.
The notion ``adiabatic condition'' here can be confusing or
even misleading. 
Because the effective field for a NFIS in the static frame
can be different from that in
 a corotating frame, the alignment/anti-alignment of a NFIS with
its effective field in one frame does not guarantee the 
alignment/anti-alignment in the other frame.
We note that the flavor evolution of a neutrino is considered to be
adiabatic if, \textit{e.g.}, it stays in the light mass eigenstate
or, equivalently, the corresponding NFIS stays aligned with its 
effective field in the \textit{static} frame.
In contrast, Eq.~\eqref{eq:pseudo-rp-cond}
is a necessary condition for all NFIS's to remain aligned or anti-aligned
with their effective fields in the appropriate \textit{corotating} frame.
As we will show at the end of
Sec.~\ref{sec:adiabaticity}, the flavor evolution
of neutrinos is not adiabatic in the conventional sense
if they stay in the regular precession mode, which satisfies
the condition in Eq.~\eqref{eq:pseudo-rp-cond}.

\subsection{Effect of the matter field%
\label{sec:matter-field}}

Ref.~\cite{Duan:2007mv} argues that,  so long as
$\sin2\thetav$ is small, ordinary matter has no effect other than
 changing the precession angular velocity $\wc$ of the system. 
This is because, if $\sin\vartheta_i\gg\sin2\thetav$,
NFIS $\sB_i$ essentially sees $\Hb_e$ as parallel to $\HV$, and
\begin{subequations}
\begin{align}
\Hb_i&=\omega_i\HV+(H_{e,x}\basev{x}+H_{e,z}\basev{z})+\mu_\nu\Sb,\\
&\simeq\omega_i^\prime\HV+\mu_\nu\Sb,
\label{eq:Hi-approx}
\end{align}
\end{subequations}
where
\begin{equation}
\omega_i^\prime\equiv\omega_i+H_{e,z}.
\end{equation}
Eq.~\eqref{eq:Hi-approx} takes a form similar to that in the e.o.m.~of
NFIS $\sB_i$ in the absence of the matter field. We note that
$\omega_i^\prime-\wc^\prime=\omega_i-\wc$, where
\begin{equation}
\wc^\prime\equiv\wc+H_{e,z},
\end{equation}
and
$\wc$ is the precession velocity in the absence of the matter field.
Therefore, Eq.~\eqref{eq:prec} would hold and the precession solution
would obtain in the corotating frame with angular velocity $\wc^\prime$
 if $H_{e,x}$ were indeed ignorable.

In the case of regular precession, 
 $\sB_i$ would be aligned 
or anti-aligned with its effective field $\Htb_i$ in the frame which
rotates around $\HV$ with angular velocity $\wc^\prime$. In this corotating
frame, the NFIS would experience an additional field $\Htb_e$ which has
magnitude $|H_{e,x}|$ and precesses around $\HV$ with angular velocity
$-\wc^\prime$. As long as
\begin{equation}
|H_{e,x}|\ll|\wc^\prime|=|\wc+H_{e,z}|,
\end{equation}
$\Htb_e$ would only introduce small perturbation in the motion of $\sB_i$ and
the regular precession solution is  a good approximation.
The regular precession approximation fails if
\begin{equation}
\wc+H_{e,z}\simeq0,
\label{eq:MSW-resonance4}
\end{equation}
which approximately 
describes the conventional MSW resonance condition 
for a neutrino with energy $E=\delta m^2/2|\wc|$.

In addition to the above effect of the matter field on the
collective precession in the regime of high neutrino number density,
the matter field can also cause
individual neutrino modes to deviate from the collective precession mode
at their MSW resonances if the neutrino number density is low. This can be
understood as follows. The effective field $\Hb_i$ for NFIS $\sB_i$ 
in the static frame
can be written as the sum of two fields: 
\begin{equation}
\Hb_\mathrm{MSW} = (\omega_i+H_{e,z}+\mu_\nu S_z)\basev{z}+ H_{e,x}\basev{x}
\end{equation}
and $\mu_\nu\Sb_\perp$,
where $\Sb_\perp$ is the  component of the total NFIS $\Sb$
perpendicular to $\basev{z}$.
As $\ntot$ and $n_e$ decrease, $\Hb_\mathrm{MSW}$ will rotate,
and, consequently, neutrino mode $i$ will encounter an MSW-like
resonance when
\begin{equation}
\omega_i+H_{e,z}+\mu_\nu S_z \simeq 0.
\end{equation}
At the same time, $\Sb_\perp$ precesses around $\HV$ 
with angular velocity $\wc$.
If the rotation speed of $\Hb_\mathrm{MSW}$ is slow enough for
the MSW resonance to be adiabatic while at the same time 
$|\mu_\nu|S_\perp\lesssim\wc$, NFIS $\sB_i$ will follow $\Hb_\mathrm{MSW}$
through the resonance and, therefore, will deviate from
 the collective precession mode.
Obviously, the lepton number $\mathcal{L}$ is not conserved
in this process.

\section{Neutrino flavor transformation in supernovae%
\label{sec:supernova}}

The problem of neutrino flavor transformation under realistic
supernova conditions is very difficult to solve. This is largely a result of
the anisotropic nature of the neutrino and antineutrino distribution
functions in the supernova environment. 
However, it has been demonstrated numerically that most
of the qualitative features of the supernova neutrino oscillation problem
 are captured by the so-called
``single-angle'' simulations (see below)
\cite{Duan:2006an,Duan:2006jv,Esteban-Pretel:2007ec}.
In this section we will show explicitly
 that the results from single-angle simulations can be
explained by the combination of MSW-like solution and the regular precession
solution discussed in the preceding sections.

\subsection{Supernova model}

In the single-angle simulations discussed in
Refs.~\cite{Duan:2006an,Duan:2006jv},
the flavor evolution of all neutrinos is assumed to be
the same as that of the neutrinos propagating along 
radial trajectories. Under this assumption, the 
neutrino-neutrino coupling strength 
is taken to be $\mu_\nu=-2\sqrt{2}\GF$
as in isotropic neutrino gases. The neutrino-neutrino intersection
angle dependence in the current-current weak interaction is
partially taken into account by introducing an effective
number density of neutrinos \cite{Fuller:1987aa,Fuller:1992aa}.
The effective 
number density $n_\nu^\eff$ of neutrinos of energy $E$ 
at radius $r$ is defined as
\begin{equation}
n_\nu^\eff(E,r) \equiv \frac{D(r/R_\nu)L_\nu f_\nu(E)}%
{2\pi R_\nu^2\langle E_\nu\rangle}.
\label{eq:n-eff}
\end{equation}
In Eq.~\eqref{eq:n-eff}, $R_\nu$ is the radius of the neutrino sphere
(here taken to be $R_\nu=11$ km),
the geometric factor
\begin{equation}
D(r/R_\nu)\equiv \frac{1}{2}
\left[1-\sqrt{1-\left(\frac{R_\nu}{r}\right)^2}\right]^2
\end{equation}
incorporates the geometric coupling and dilution of
anisotropic neutrino beams as a function of radius $r$,
 $L_\nu$ and $\langle E_\nu\rangle$ are the
luminosity and average energy of the neutrino species, respectively,
and $f_\nu(E)$
is the (normalized) energy distribution function for neutrinos.

In the single-angle simulations, luminosities for all neutrino 
species are taken to be the same, and two typical late-time
supernova neutrino luminosity values
$L_\nu=5\times10^{51}$ erg/s and $10^{51}$ erg/s have been used.
The energy distribution function $f_\nu(E)$ for neutrinos is taken
to be of the Fermi-Dirac form with two parameters $(T_\nu, \eta_\nu)$,
\begin{equation}
f_\nu (E) \equiv \frac{1}{F_2(\eta_\nu)} \frac{1}{T_\nu^3}
\frac{E^2}{\exp(E/T_\nu-\eta_\nu)+1},
\label{eq:energy-dist}
\end{equation}
where we take the degeneracy parameter to be $\eta_\nu=3$, $T_\nu$ is the
 neutrino temperature, and
\begin{equation}
F_k (\eta) \equiv \int_0^\infty \frac{x^k \,\ud x}{\exp(x-\eta)+1}.
\end{equation}
The values of $T_\nu$ for various neutrino species are determined
from our chosen average energies:
 $\langle E_{\nu_e}\rangle = 11$ MeV, 
$\langle E_{\bar\nu_e} \rangle= 16$ MeV, and
$\langle E_{\nu_\mu} \rangle = \langle E_{\bar\nu_\mu} \rangle= 
\langle E_{\nu_\tau}\rangle = \langle E_{\bar\nu_\tau} \rangle = 25$ MeV
\cite{Qian:1993dg}.

The simulations use
a simple density profile. The net electron density
at radii sufficiently above the neutrino sphere
 is taken to be \cite{Fuller:2005ae}
\begin{equation}
n_e=Y_e \frac{2\pi^2}{45} g_\mathrm{s} 
\left(\frac{M_\mathrm{NS}\, m_N}{m_\mathrm{Pl}^2}\right)^3 S^{-4} r^{-3},
\label{eq:ne}
\end{equation}
where $Y_e$ is the electron fraction (here taken to be $Y_e=0.4$), 
$g_\mathrm{s}=11/2$ is the statistical
weight in relativistic particles, $M_\mathrm{NS}=1.4M_\odot$ is the mass
of the neutron star, $m_N$ is the mass of the nucleon, $m_\mathrm{Pl}$
is the Plank mass, and $S=140$ is the entropy in units
of Boltzmann's constant per baryon. 

The matter density near the neutrino sphere
is  much larger than what Eq.~\eqref{eq:ne} calculates.
The simulations discussed in Refs.~\cite{Duan:2006an,Duan:2006jv}
have adopted an exponential density profile in the region near the neutrino
sphere.
We note that a large $n_e$ will keep neutrinos in their
initial flavor eigenstates. For the MSW-like solution,
the $n_e$ profile in Eq.~\eqref{eq:ne} is large enough to
keep neutrinos in their initial
flavor states to  $r\gg R_\nu$,
and the exponential density profile near the neutrino sphere will
not affect our  analysis below.

\subsection{Comparison of numerical and analytical solutions%
\label{sec:comparison}}

For $2\times2$ flavor mixing, neutrinos starting in pure $\nu_e$
($\bar\nu_e$) and $\nu_\tau$ ($\bar\nu_\tau$) states with the same energy
evolve in the same way and can be viewed as the same neutrino mode.
This is because NFIS $\sB_i$ and $-\sB_i$ follow the same
e.o.m. As  all neutrinos are assumed to be in pure
flavor eigenstates at the neutrino sphere, a neutrino mode $i$ is
uniquely designated by $\omega_i$ [defined in Eq.~\eqref{eq:muV-def}],
and we can take $\sum_i\rightarrow\int_{-\infty}^{\infty}\ud\omega$.

We define the total neutrino number density at radius $r$ as
\begin{equation}
\begin{split}
\ntot(r) &= \int_0^\infty |n_{\nu_e}^\eff(E,r)-n_{\nu_\tau}^\eff(E,r)|\,\ud E\\
&\quad+ \int_0^\infty |n_{\bar\nu_e}^\eff(E,r)-n_{\bar\nu_\tau}^\eff(E,r)|\,\ud E.
\end{split}
\label{eq:ntot}
\end{equation}
Note that $\ntot$ defined above takes advantage of the
equivalent flavor evolution of $\nu_e$ ($\bar\nu_e$) and $\nu_\tau$ 
($\bar\nu_\tau$), and is not the simple sum of
number densities of all neutrinos. We define the NFIS mode
distribution function as
\begin{equation}
\ft_\omega = \frac{|\ud E/\ud \omega|}{\ntot(R_\nu)} \times\left\{
\begin{array}{ll}
|n_{\nu_e}^\eff(E,R_\nu)-n_{\nu_\tau}^\eff(E,R_\nu)|
&\text{if }\omega>0,\\
|n_{\bar\nu_e}^\eff(E,R_\nu)-n_{\bar\nu_\tau}^\eff(E,R_\nu)|
&\text{if }\omega<0.
\end{array}
\right.
\label{eq:ft}
\end{equation}
Because $\mu_\nu<0$ and $\nu_e$'s are dominant in number in supernovae,
the vector $\mu_\nu\Sb$ is initially in the direction opposite to  $\basef{z}$.
Taking this into account, we have
\begin{equation}
\frac{\Hb_\omega}{H_\omega}\simeq
\frac{\Hb_e+\mu_\nu\Sb}{H_\omega}\simeq-\basef{z}
\label{eq:Homega-dir}
\end{equation}
at the neutrino sphere, where $\Hb_\omega$ is the total
effective field for neutrino mode  $\omega$. 
Eq.~\eqref{eq:Homega-dir} is true for any neutrino mode $\omega$.

Noting that $\nu_e$/$\bar\nu_\tau$ and
$\bar\nu_e$/$\nu_\tau$ are represented by NFIS's aligned and anti-aligned
with $\basef{z}$, respectively, we find the alignment factor $\epsilon_\omega$
for neutrino mode $\omega$ to be
\begin{equation}
\epsilon_\omega = \left\{
\begin{array}{ll}
-\sgn (n_{\nu_e}^\eff(E,R_\nu)-n_{\nu_\tau}^\eff(E,R_\nu))
&\text{if }\omega>0,\\
\sgn (n_{\bar\nu_e}^\eff(E,R_\nu)-n_{\bar\nu_\tau}^\eff(E,R_\nu))
&\text{if }\omega<0,
\end{array}
\right.
\label{eq:epsilon}
\end{equation}
where $\sgn(\xi)\equiv\xi/|\xi|$ is the sign of $\xi$.

\begin{figure*}
\begin{center}
$\begin{array}{@{}c@{\hspace{\myfigsep}}c@{}}
\includegraphics*[width=\myfigwid, keepaspectratio]{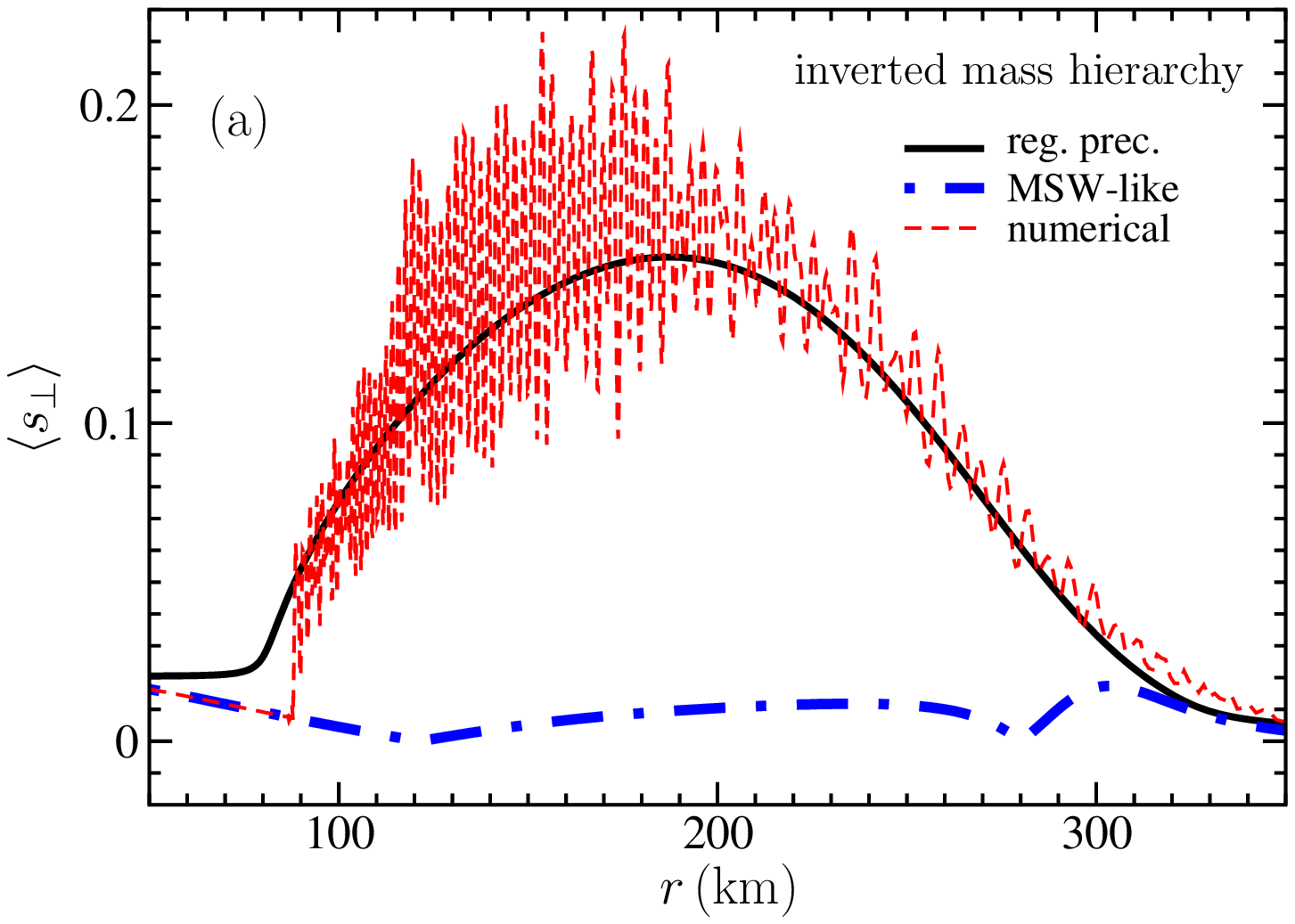} &
\includegraphics*[width=\myfigwid, keepaspectratio]{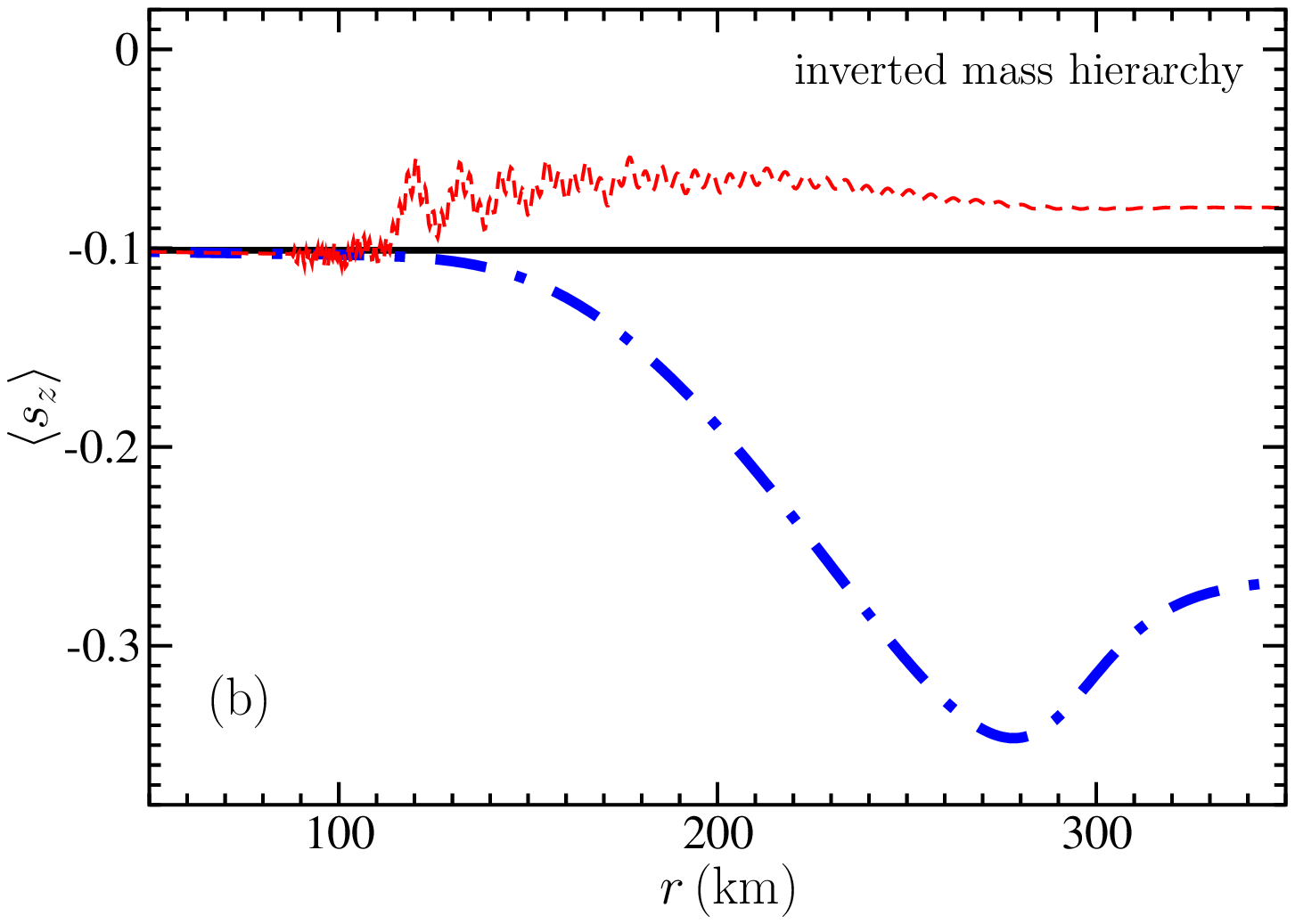} \\
\includegraphics*[width=\myfigwid, keepaspectratio]{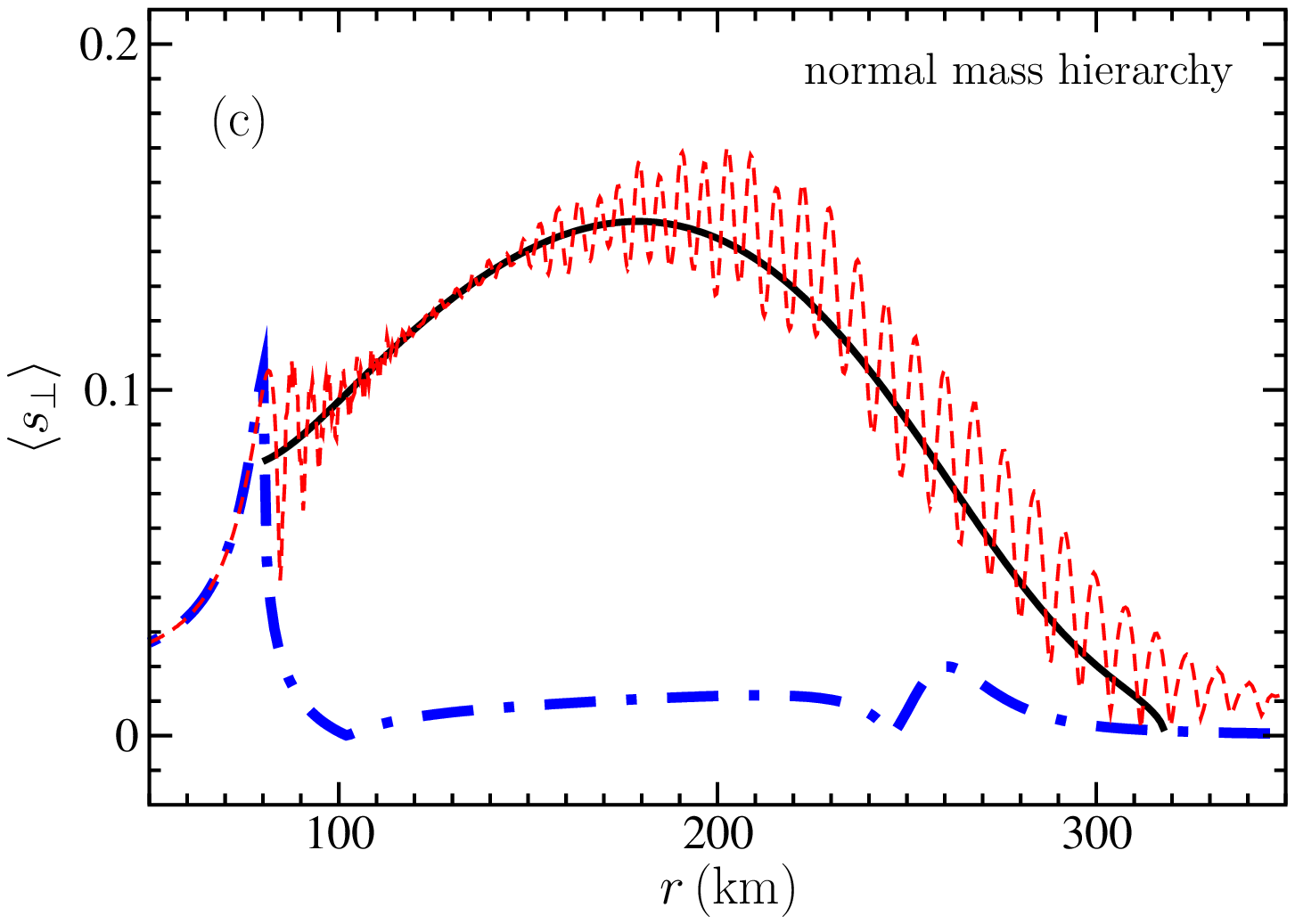} &
\includegraphics*[width=\myfigwid, keepaspectratio]{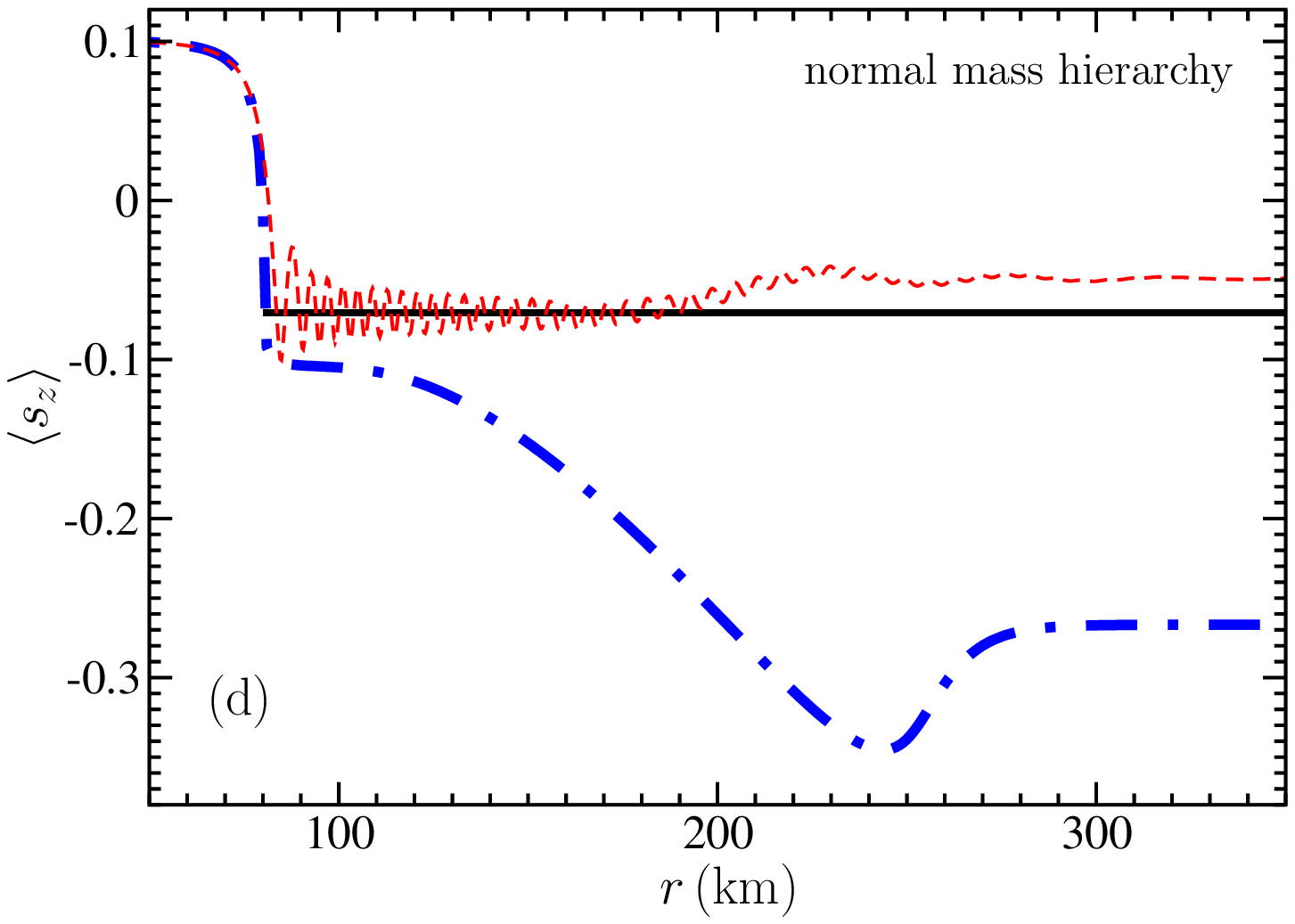}
\end{array}$
\end{center}
\caption{\label{fig:s-r-L551}(Color online)
Plots of $\langle s_\perp(r)\rangle$ (left panels) 
and $\langle s_z(r)\rangle$ (right panels) for both the inverted (upper panels)
and normal (lower panels) neutrino mass hierarchies.
The dashed lines show numerical simulation results from
 Ref.~\cite{Duan:2006an}. 
The solid lines show the regular precession solution.
The dot-dashed lines show the  MSW-like solution.
The  luminosity for all neutrino species
is taken to be $L_\nu=5\times10^{51}$ erg/s.}
\end{figure*}     

\begin{figure*}
\begin{center}
$\begin{array}{@{}c@{\hspace{\myfigsep}}c@{}}
\includegraphics*[width=\myfigwid, keepaspectratio]{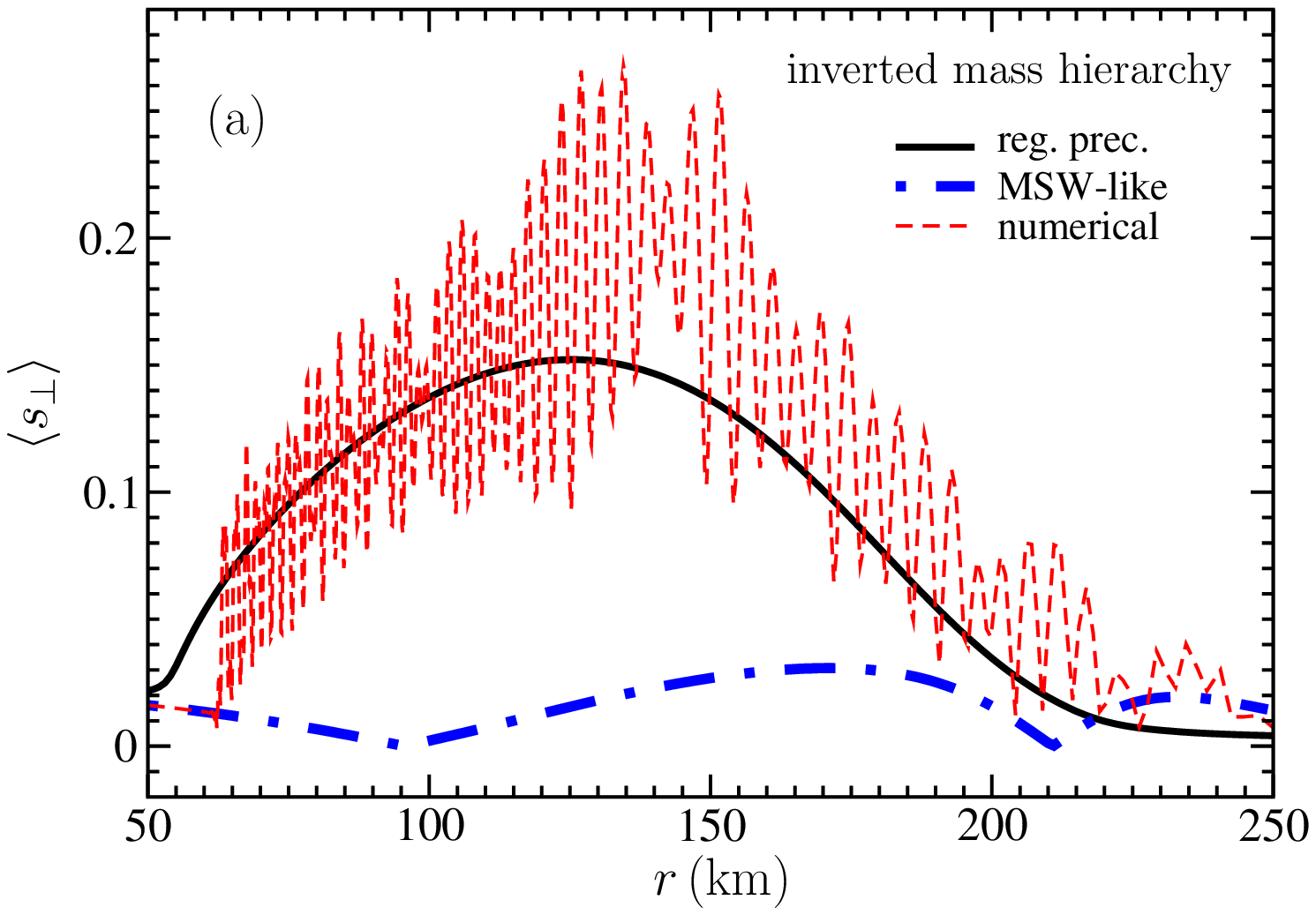} &
\includegraphics*[width=\myfigwid, keepaspectratio]{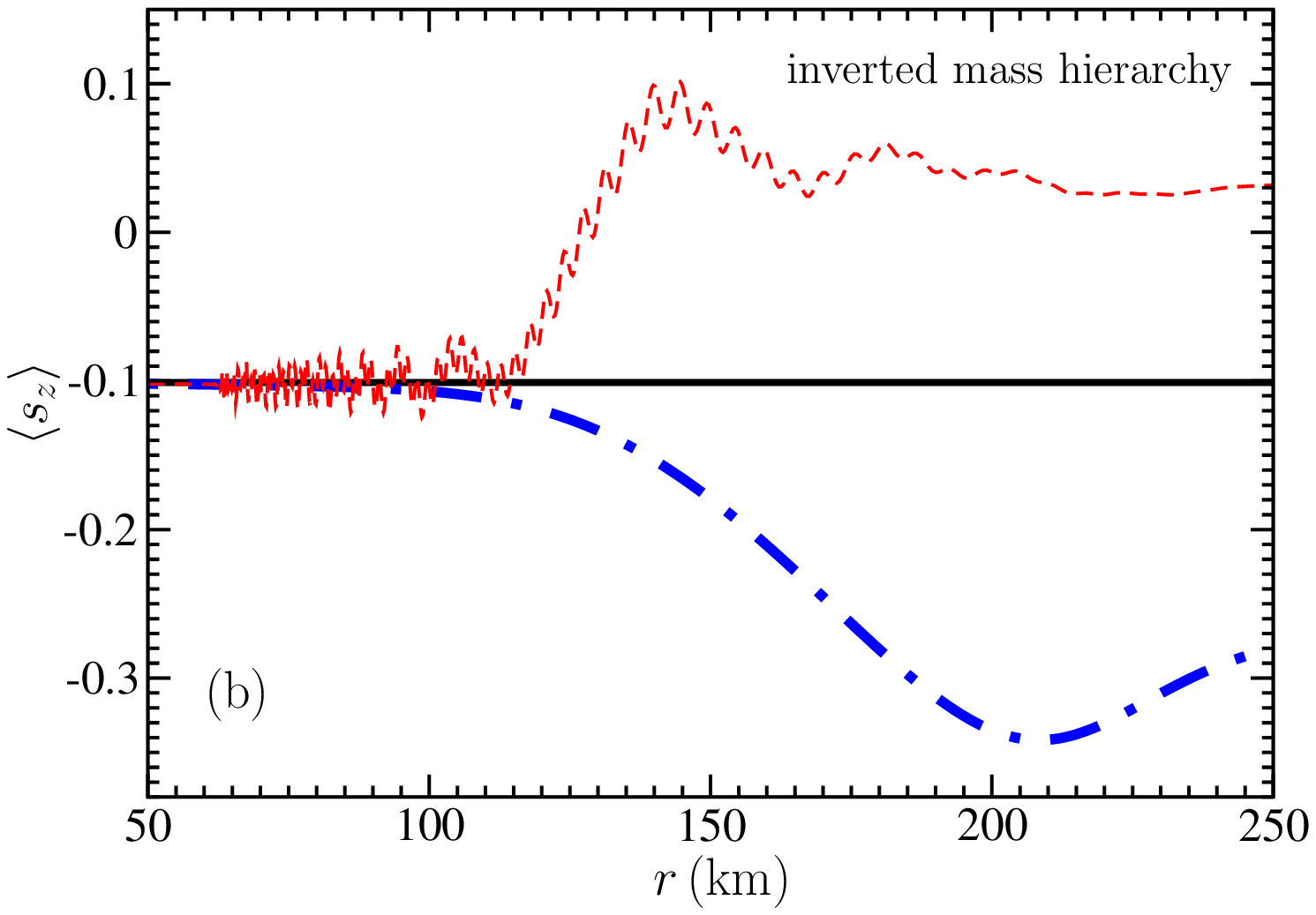} \\
\includegraphics*[width=\myfigwid, keepaspectratio]{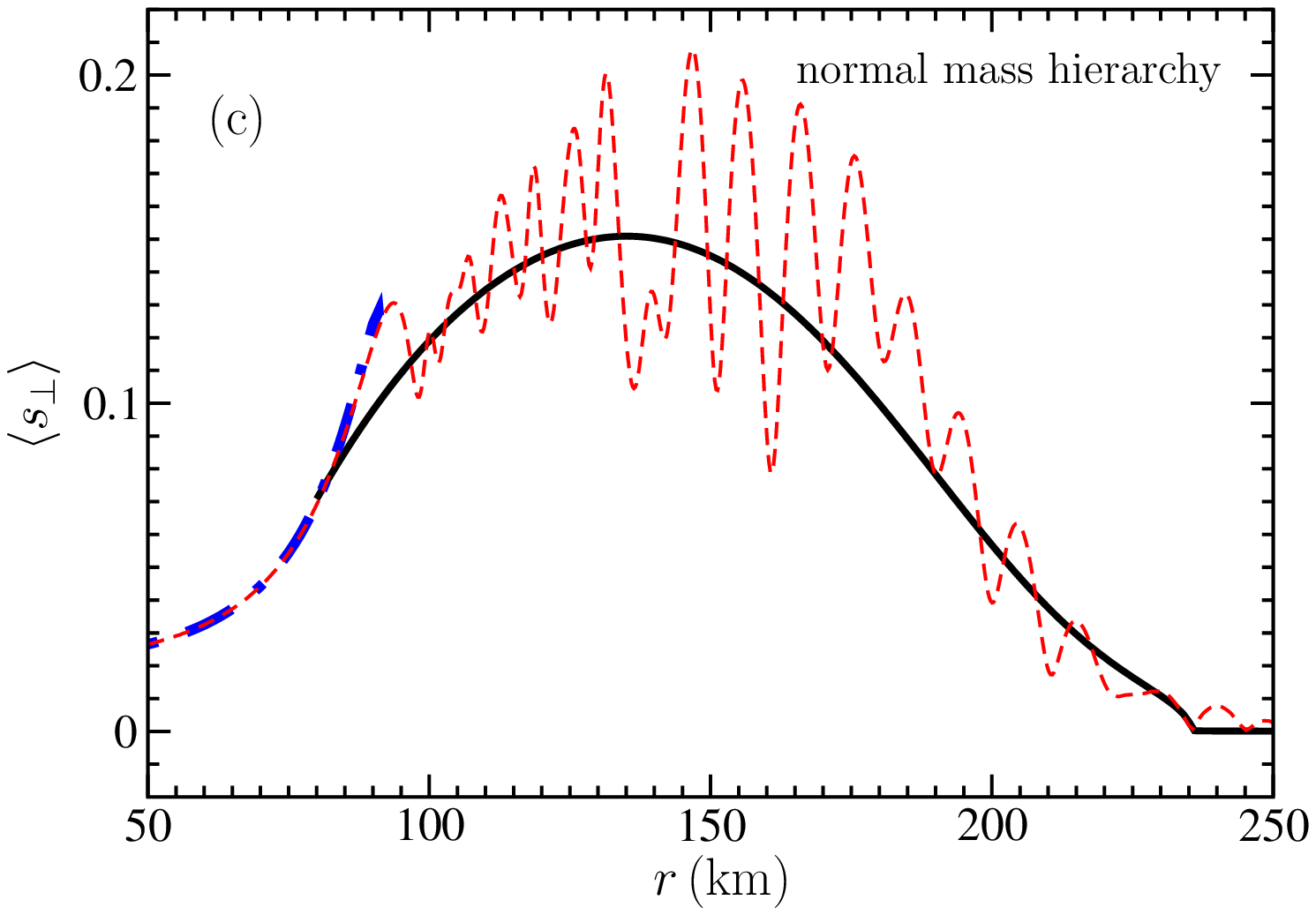} &
\includegraphics*[width=\myfigwid, keepaspectratio]{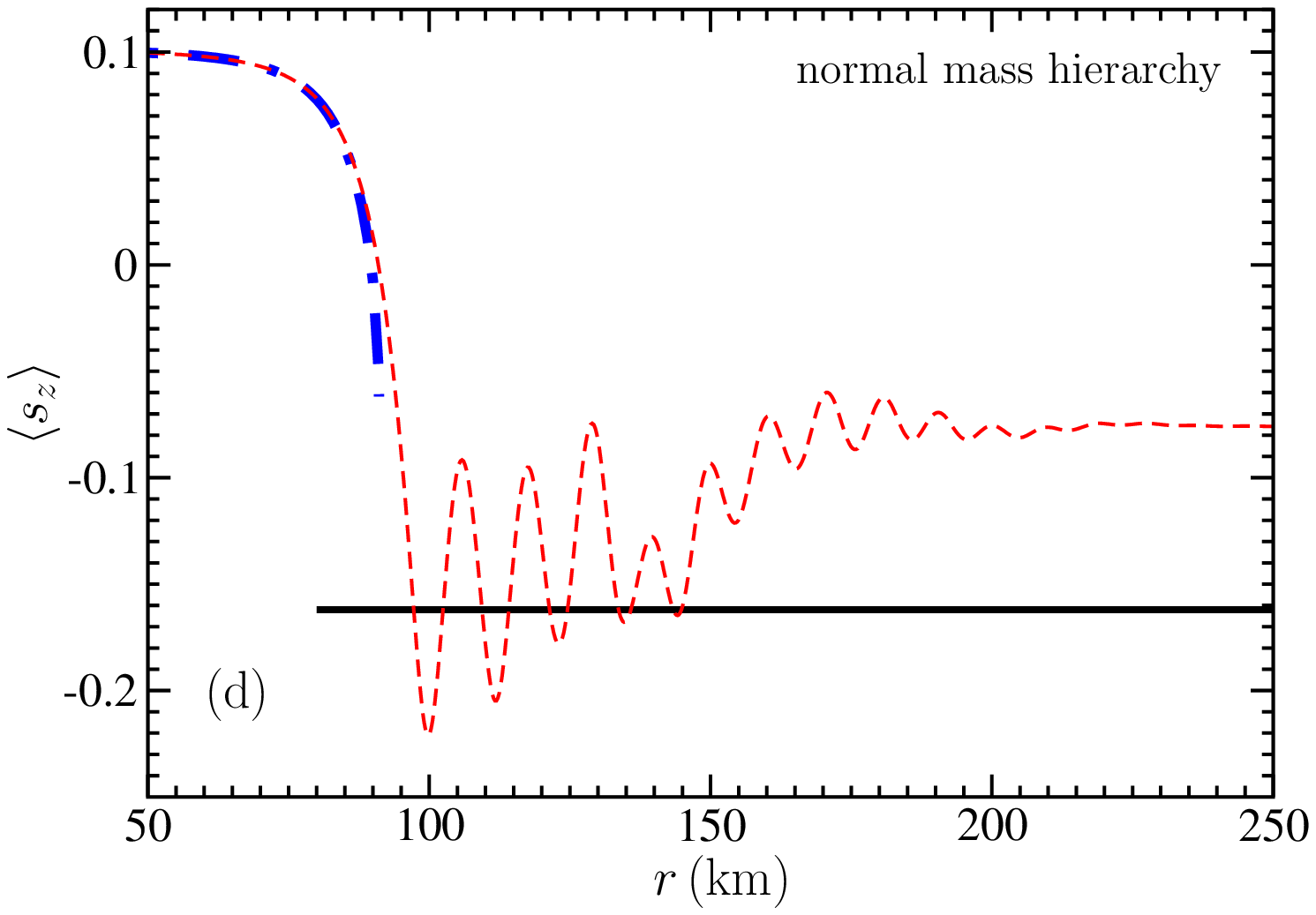}
\end{array}$
\end{center}
\caption{\label{fig:s-r-L51}(Color online)
Same as Fig.~\ref{fig:s-r-L551} except that 
the luminosity for all neutrino species is taken to be $L_\nu=10^{51}$ erg/s
in this figure.}
\end{figure*}     

Using Eqs.~\eqref{eq:Sxz}, \eqref{eq:ntot}, \eqref{eq:ft} and \eqref{eq:epsilon}
we are able to find the MSW-like solution in the same supernova model 
as adopted in
the single-angle simulations. Taking 
$\delta m^2=3\times10^{-3}\,\mathrm{eV}^2$
(close to the value associated with atmospheric neutrino oscillations),
$\thetav=0.1$ (normal neutrino mass hierarchy)
or $\pi/2-0.1$ (inverted neutrino mass hierarchy),
$L_\nu=5\times10^{51}$ erg/s,
we compute the average NFIS
\begin{equation}
\langle\sB\rangle \equiv \frac{\Sb}{\ntot}
\end{equation}
in the MSW-like solution. We plot $\langle s_\perp\rangle=|\langle s_x\rangle|$
and $\langle s_z\rangle$, the perpendicular and $z$ components,
respectively, of
the average NFIS in the vacuum mass basis, as dot-dashed lines
in Fig.~\ref{fig:s-r-L551}.

We also obtain the regular precession solution
under the same conditions. For the inverted mass hierarchy case, 
we determine the value of
$\langle s_z\rangle$ in the regular precession solution
 using  the lepton number 
${\cal{L}}=2\langle s_z\rangle=2S_z/n_\nu^\mathrm{tot}$ 
at the neutrino sphere [see discussions around Eq.~\eqref{eq:sz-invt}].
For the normal mass hierarchy case, 
we determine the value of
$\langle s_z\rangle$ using the transition energy $\EC$
[see discussions around
Eq.~\eqref{eq:sz-swap}]. 
Note that the swapping of energy spectra above/below $\EC$
(stepwise swapping) is a natural result of the regular precession
solution.
With $S_z=\ntot s_z$ we solve Eqs.~\eqref{eq:unity}
and \eqref{eq:wc} for  $S_x$ (or $S_\perp$ in the static frame) and 
$\omega_\mathrm{pr}$.  The values of 
$\langle s_\perp\rangle= S_\perp/\ntot$ and
$\langle s_z\rangle$ corresponding to this solution for both the inverted
and normal mass hierarchies are plotted as solid lines in 
Fig.~\ref{fig:s-r-L551}.

 For comparison with the MSW-like and 
regular precession solutions, we extract the values of
$\langle s_\perp\rangle$ and $\langle s_z\rangle$ in the single-angle
simulations presented in Ref.~\cite{Duan:2006an}. These numerical results are
plotted as dashed lines in Fig.~\ref{fig:s-r-L551}.

For the inverted mass hierarchy case 
[Fig.~\ref{fig:s-r-L551}(a--b)] we observe that $\langle s_\perp\rangle$
in the numerical simulation follows the MSW-like solution
until radius $r_\mathrm{X}\simeq88$ km, but  then abruptly  jumps
out of the MSW-like solution. Thereafter, $\langle s_\perp\rangle$ 
 roughly follows the regular
precession solution. The reason for this change in
the behavior of the NFIS's is discussed extensively in 
Ref.~\cite{Duan:2007mv}. 
Briefly speaking, supernova neutrinos
naturally form a ``bipolar system'' with the neutrinos forming
 two groups of NFIS's pointing in  opposite directions 
in flavor space \cite{Duan:2006an}.
This bipolar system is roughly equivalent to a gyroscopic
pendulum in flavor space \cite{Hannestad:2006nj}. For the
inverted mass hierarchy case, the gyroscopic pendulum
is initially in the highest position, and $(\ntot)^{-1}$
plays the role of (variable) gravity. For a given value of internal spin
(determined by $\nu_e$--$\bar\nu_e$ asymmetry), the gyroscopic pendulum
will not fall until the ``gravity'' is large enough and $\ntot$
drops below some critical value $n_\nu^\mathrm{c}$. If the gravity
increases very slowly, so that the symmetry around 
$\HV$ is preserved, the gyroscopic pendulum
will precess steadily. This corresponds to the regular precession
solution. However, the symmetry is usually broken near the point where 
$\ntot(r)\simeq n_\nu^\mathrm{c}$. Below this critical density,
 the pendulum will develop
significant nutation in addition to the precession.

We note that the fast oscillations of $\langle s_\perp\rangle$
around the regular precession solution correspond to the nutation mode.
We also note that $\langle s_z\rangle$ roughly stays constant in the region
 $r\lesssim 115$ km and is
\begin{subequations}
\label{eq:sz-invt}
\begin{align}
\langle s_z\rangle &\equiv \frac{\Sb\cdot\HV}{\ntot},\\
&\simeq -\frac{\Sb\cdot\basef{z}}{\ntot} 
= \frac{1}{2}\int_{-\infty}^\infty \epsilon_\omega\ft_\omega\ud\omega,\\
&\simeq -0.1.
\end{align}
\end{subequations}
This is because both the collective precession
and nutation modes conserve the lepton number 
$\mathcal{L}=2\langle s_z\rangle$.
The value of $\langle s_z\rangle$ is subsequently changed because
of matter effects,  which do not conserve $\mathcal{L}$.
However, $\langle s_\perp\rangle$ in the numerical solution
still roughly follows that of the regular precession solution, even
after $\langle s_z\rangle$ starts to evolve. This seems to imply
that the neutrino system is still described by the regular
precession solution, except for the neutrino modes which have
dropped out of the collective precession.

For the normal mass hierarchy case 
[Fig.~\ref{fig:s-r-L551}(c--d)] we observe that $\langle s_\perp\rangle$
and $\langle s_z\rangle$ in the numerical simulation follow the MSW-like 
solution until radius $r\simeq83$ km. At this point, $\langle \sB\rangle$
has almost completely flipped its direction. This corresponds to
the complete flavor transformation in both $\nu_e\leftrightharpoons\nu_\tau$
and $\bar\nu_e\leftrightharpoons\bar\nu_\tau$
channels. In the flavor pendulum analogy,
the gyroscopic pendulum has been raised from the lowest position to
the highest position, and one expects it to fall when
$\ntot(r)\lesssim n_\nu^\mathrm{c}$. Indeed, 
for large enough radii,
both $\langle s_\perp\rangle$
and $\langle s_z\rangle$ leave the MSW-like solution and
start to oscillate around the regular precession solution with
$\langle s_z\rangle\simeq -0.07$.

We have also obtained the three kinds of solutions for a smaller neutrino
luminosity, $L_\nu=10^{51}\,\mathrm{erg/s}$, with other parameters unchanged.
These results are plotted in Fig.~\ref{fig:s-r-L51}.
Comparing Fig.~\ref{fig:s-r-L551} and \ref{fig:s-r-L51}, one can see that
the small luminosity cases are very similar to their
 large luminosity counterparts.
However, we note that, for the inverted mass hierarchy case,
$\langle s_\perp\rangle$  leaves
the MSW-like solution at $r_\mathrm{X}\simeq 63$ km, which is smaller than
the $r_\mathrm{X}$ value in the large luminosity case. This is because with
smaller $L_\nu$ the total
neutrino number density $\ntot$ reaches $n_\nu^\mathrm{c}$ earlier.
 For the normal mass hierarchy case, our calculation procedure
cannot find the MSW-like solution in a small radius
interval immediately beyond $r\simeq91$ km.
We have taken $\langle s_z\rangle\simeq -0.16$ for the regular precession
solution in this case.
We also note that the value of $\langle s_z\rangle$ changes
significantly in both small luminosity cases. This is because
with smaller neutrino number densities,
the matter field drives more neutrinos or antineutrinos off the track
of the collective
precession solution.

\subsection{Stepwise swapping in neutrino energy spectra%
\label{sec:swapping}}

As we have seen in Sec.~\ref{sec:comparison}, supernova neutrinos
evolve initially according
to the MSW-like solution, but  subsequently roughly follow
the regular precession solution. In the latter
solution, all NFIS's are aligned or anti-aligned with effective
fields in the appropriate corotating frame (see Sec.~\ref{sec:prec-sol-der}).
It was first suggested in  Ref.~\cite{Duan:2006an} and further
explained in Ref.~\cite{Duan:2007mv} that, if this alignment in
the corotating frame is maintained, $\nu_e$ and $\nu_\tau$
will swap their energy spectra for energy smaller (larger) than
a transition energy
$E_\mathrm{C}$ in the normal (inverted) mass hierarchy case.
This phenomenon is also referred to as a ``spectral split'' and
the transition energy is determined by 
the conservation of lepton
number $\mathcal{L}$ \cite{Raffelt:2007cb}. 
This stepwise swapping of neutrino energy spectra
has been observed in the numerical simulations presented in
Refs.~\cite{Duan:2006an,Duan:2006jv}. 

The phenomenon of
the stepwise spectrum swapping (or spectral split)
can be understood as follows.
In the regular precession solution, NFIS $\sB_\omega$
stays aligned or anti-aligned with $\Hb_\omega$ and
 $\sB_\omega=\epsilon_\omega\Htb_\omega/2\tilde{H}_\omega$. 
The effective field $\Htb_\omega$ in the corotating frame
becomes
\begin{equation}
\Htb_\omega=(\omega-\wc^0)\HV
\end{equation}
as $\ntot\rightarrow0$,
where $\wc^0$ is the value of $\wc$ at $\ntot=0$. Therefore, one has
\begin{equation}
(s_{\omega,z})_{\ntot=0}=\frac{\epsilon_\omega}{2}\sgn(\omega-\wc^0).
\label{eq:sz-final}
\end{equation}
In other words, the spectrum of $\epsilon_\omega s_{\omega,z}$
is split into two parts  when $\ntot\rightarrow0$: 
the lower part  ($\omega<\wc^0$) takes the value $-1/2$ while
the upper part ($\omega>\wc^0$) takes the value $+1/2$.

For $\sin2\thetav\ll1$ the NFIS for mode $\omega$
in the limit of $\ntot=0$  is
\begin{subequations}
\begin{align}
(\sB_\omega)_{\ntot=0}
&= \HV(s_{\omega,z})_{\ntot=0},\\
&\simeq \frac{\epsilon_\omega}{2}\Xi\,\sgn(\omega-\wc^0)
\basef{z},
\label{eq:sstep}
\end{align}
\end{subequations}
where $\Xi=+1$ ($-1$) for the normal (inverted) mass hierarchy.
Eq.~\eqref{eq:sstep} means that
 supernova neutrinos  are almost in pure flavor eigenstates
when $\ntot\rightarrow0$. We note that neutrinos are in flavor
eigenstates at the neutrino sphere and
\begin{equation}
(\sB_\omega)_{r=R_\nu}=-\frac{\epsilon_\omega}{2}\basef{z}.
\end{equation}
Therefore, the probability for a neutrino mode $\omega$
to stay in its original state is
\begin{subequations}
\begin{align}
P_{\nu\nu}(\omega)&=\frac{1}{2}+2(\sB_\omega)_{r=R_\nu}\cdot(\sB_\omega)_{\ntot=0},\\
&\simeq \frac{1}{2}
\left[1-\Xi\,
\sgn(\omega-\wc^0)\right].
\label{eq:step}
\end{align}
\end{subequations}
According to Eq.~\eqref{eq:step}, a stepwise swapping can
occur in the final neutrino energy spectra. For
$\wc^0>0$ and $\thetav\simeq\pi/2$ (inverted mass hierarchy), 
$\nu_e$ and $\nu_\tau$
will swap their energy spectra for energies above
\begin{equation}
E_\mathrm{C} = \frac{\delta m^2}{2 \wc^0}.
\end{equation}
For $\wc^0>0$ and $\thetav\ll1$ (normal mass hierarchy), $\nu_e$ and $\nu_\tau$
will swap their energy spectra for energies below $E_\mathrm{C}$.

Assuming stepwise spectrum swapping, the value of $\wc^0$
can be determined from $\langle s_z\rangle$ without finding the
complete regular precession solution \cite{Raffelt:2007cb}:
\begin{equation}
2\langle s_z\rangle= \int_{\wc^0}^\infty\epsilon_\omega\ft_\omega\ud\omega
-\int_{-\infty}^{\wc^0}\epsilon_\omega\ft_\omega\ud\omega.
\label{eq:sz-swap}
\end{equation}
Using the values of $\langle s_z\rangle$ in the regular precession
solutions [Eq.~\eqref{eq:sz-invt}], we obtain $E_\mathrm{C}\simeq8.4$
MeV for the inverted mass hierarchy cases. This value agrees well
with the numerical simulation results 
(see Fig.~9(c) in Ref.~\cite{Duan:2006an}).
For the normal
mass hierarchy scenarios, we actually
obtain $\langle s_z\rangle$ in the regular precession solution
by using Eq.~\eqref{eq:sz-swap} and
the values of $\EC$ in the numerical results
(Fig.~9(a) in Ref.~\cite{Duan:2006an}) which are
approximately $7.8$ MeV and $9.5$ MeV
for the large and small neutrino luminosity cases, respectively. 
As shown in Figs.~\ref{fig:s-r-L551}(c) and \ref{fig:s-r-L51}(c), 
the precession
solutions determined by these values of $\EC$ agree well
with the numerical simulation results.
For the parameters we have chosen, the stepwise swapping  occurs only
in the neutrino sector. This is because $\nu_e$ is the dominant
(in terms of number flux)
neutrino species in supernovae.

\subsection{Adiabaticity%
\label{sec:adiabaticity}}

The flavor evolution of neutrino mode $\omega$ is adiabatic so long as
the angle between $\sB_\omega$ and $\Hb_\omega$ remains constant. 
It can be shown that the flavor evolution of neutrino mode $\omega$  
is adiabatic if $\Hb_\omega$ rotates at a rate much slower than $H_\omega$,
 or, in other words,
\begin{equation}
\left|\Hb_\omega\times\frac{\ud \Hb_\omega}{\ud t}\right| \ll H_\omega^3
\label{eq:adiabatic-strong}
\end{equation}
(see, \textit{e.g.}, Ref.~\cite{Duan:2005cp}).
Adiabaticity is guaranteed if the condition in Eq.~\eqref{eq:adiabatic-strong}
is satisfied. However, when neutrino densities are large, the system
can evolve adiabatically even if the condition in  
Eq.~\eqref{eq:adiabatic-strong} is \textit{not} satisfied.

If neutrinos experience collective
flavor transformation,  the total NFIS $\Sb$, and
therefore, effective field
$\Hb_\omega=\omega\HV+\Hb_e+\mu_\nu\Sb$
move in correlation with the motion of individual NFIS $\sB_\omega$.
In this case, $\Hb_\omega$ can move very fast, with $\sB_\omega$
staying aligned or anti-aligned with $\Hb_\omega$, and the flavor
evolution is, therefore,  still adiabatic.
We expect this  to be the case as neutrinos transition
 from
the MSW-like solution through the collective nutation mode
to the regular precession solution.
 As a result, the alignment factor $\epsilon_\omega$
for neutrino mode $\omega$ remains constant during the transition. This
was implicitly assumed in the discussion on stepwise spectrum swapping 
in Sec.~\ref{sec:swapping}.

On the other hand, we note that  flavor evolution through
the regular precession solution is \textit{not} adiabatic.
Although the pseudo-regular precession condition in
Eq.~\eqref{eq:pseudo-rp-cond} is in a form similar
to the adiabatic condition in Eq.~\eqref{eq:adiabatic-strong},
it simply guarantees that NFIS $\sB_\omega$ remains aligned or anti-aligned
with the effective field $\Htb_\omega$ in the corotating frame,
 but \textit{not} with
$\Hb_\omega$ in the static frame. Therefore, neutrino mode $\omega$
does not stay in the same mass eigenstate throughout its evolution.
One may consider two neutrino modes $\wc^0+\Delta\omega$ and 
$\wc^0-\Delta\omega$ which are initially in pure $\nu_e$ states.
If the flavor evolution is adiabatic for both neutrinos,
both NFIS's should stay anti-aligned with their effective fields 
in the static frame
throughout the evolution. However, we know that this is not the case.
Because of the stepwise swapping in neutrino energy spectra,
one of the two NFIS's will become aligned with its effective
field in the static frame
when $\ntot\rightarrow0$. Therefore, the flavor evolution
of neutrinos in the regular precession solution cannot be adiabatic
for all neutrinos.

\section{Conclusions%
\label{sec:conclusions}}

We have derived a pair of equations from which
it is possible to find
 a quasi-static MSW-like solution
for an isotropic neutrino gas  with specified electron
 and neutrino number densities. This solution is
the natural extension of the conventional MSW solution, but includes
neutrino self-coupling. We have shown that the condition for this MSW-like
solution to be adiabatic  is a necessary condition for neutrinos
to follow such a solution.
We have also discussed the previously discovered
regular precession solution and the scenarios where it may break down
in the presence of ordinary matter.

We have compared the results of the detailed simulations in
Refs.~\cite{Duan:2006an,Duan:2006jv} with the corresponding MSW-like
and regular precession solutions.
 This comparison
clearly shows that many features in the numerical simulations,
including the stepwise swapping in neutrino energy spectra,
can be explained by combinations of these two analytical solutions.
We have also discussed the adiabaticity of flavor evolution in dense
neutrino gases.

We  emphasize that the discovery of the MSW-like solution and
the regular precession solution by no means obviates the need for further
numerical simulations. For example, neutrinos only roughly follow
the regular precession solution,
typically exhibiting significant nutation mode behavior. 
In fact, the collective
nutation mode is responsible for driving neutrinos away from the MSW-like
solution and towards the regular precession solution. We note
that so far the only method to quantitatively 
follow neutrino flavor transformation
with the nutation mode  is through numerical simulations.
We also note that, in the normal mass hierarchy scenario, 
the  value of the lepton number, a key parameter for
 the regular precession solution, is determined numerically.
Nevertheless, the combination of
the MSW-like solution and the regular precession solution
offers a way to gain key insights into the results obtained from
numerical simulations.

\begin{acknowledgments}
We would like to acknowledge valuable discussions with J.~Carlson
and Jun Hidaka.
This work was supported in part by 
NSF grant PHY-04-00359 and
the TSI collaboration's DOE SciDAC grant at UCSD, and
DOE grant DE-FG02-87ER40328 at UMN.
\end{acknowledgments}

\bibliography{ref}

\end{document}